\documentclass[twocolumn]{aastex7}

\usepackage{amsmath}

\begin{document}

\title{Testing the isothermal Jeans model for self-interacting dark matter halos in the collapse phase}

\author[0009-0004-0904-7400]{Shubo Li}
\affiliation{School of Physics and Astronomy, Beijing Normal University, Beijing 100875, China}
\affiliation{National Astronomical Observatories, Chinese Academy of Sciences, 20A Datun Road, Chaoyang District, Beijing 100101, China}
\affiliation{School of Astronomy and Space Science, University of Chinese Academy of Sciences, Beijing 100049, China}
\email[show]{lishubo19@mails.ucas.ac.cn}

\author[orcid=0000-0002-6619-4480,gname=Moritz S.,sname=Fischer]{Moritz S.\ Fischer}
\affiliation{Donostia International Physics Center (DIPC), Paseo Manuel de Lardizabal 4, 20018 Donostia-San Sebastián, Spain}
\email[show]{moritz.fischer@dipc.org}

\author[0009-0008-2362-1367]{Zixiang Jia}
\affiliation{Department of Astronomy, Peking University, Beijing 100871, China}
\email{jiazixiang@stu.pku.edu.cn}

\author[0000-0001-6115-0633]{Fangzhou Jiang}
\affiliation{Kavli Institute for Astronomy and Astrophysics, Peking University, Beijing 100871, China}
\email{fangzhou.jiang@pku.edu.cn}

\author[0000-0003-3899-0612]{Ran Li}
\affiliation{School of Physics and Astronomy, Beijing Normal University, Beijing 100875, China}
\affiliation{School of Astronomy and Space Science, University of Chinese Academy of Sciences, Beijing 100049, China} 
\email{liran@bnu.edu.cn}

\author[0000-0002-8421-8597]{Hai-Bo Yu}
\affiliation{Department of Physics and Astronomy, University of California, Riverside, CA 92521, USA}
\email{hai-bo.yu@ucr.edu}

\begin{abstract}
We benchmark the semianalytical isothermal Jeans model against a high-resolution isolated $N$-body simulation that follows a self-interacting dark matter (SIDM) halo into deep core collapse. The model accurately reproduces the density evolution through much of the collapse phase, although it does not capture the sharp rise in central velocity dispersion during collapse. When applied to strong gravitational lensing observables, such as the projected mass and logarithmic density slope of SIDM halos, the Jeans model tracks the simulated evolution more closely than the parametric approach in the deep collapse regime. Our results demonstrate that the isothermal Jeans model provides a reliable and computationally efficient description of SIDM halo evolution.
\end{abstract}

\keywords{\uat{Dark matter}{353}, \uat{Dark matter halos}{1880}}

\section{Introduction} 
The particle nature of dark matter remains an outstanding question in cosmology. While the standard cold dark matter (CDM) model has been remarkably successful on large scales \citep[e.g.,][]{PlanckCollaboration2020}, it faces persistent challenges on galactic scales \citep[for a review of small-scale challenges, see][]{Bullock2017, Sales2022}. Self-interacting dark matter (SIDM) has emerged as a compelling alternative to address these discrepancies \citep{Spergel2000, Review2018, Review2025}. 

In SIDM models, nongravitational interactions among dark matter particles facilitate the transport of heat within dark matter halos. This heat transport initially thermalizes the halo center, creating a constant-density core. In the subsequent stage, heat continues to flow outward due to the temperature difference between the thermalized core and the cooler halo outskirts. As the core loses energy, it contracts under self-gravity, and the negative specific heat capacity of self-gravitating systems causes this contraction to be accompanied by an increase in the core temperature. The resulting steepening of the velocity-dispersion gradient strengthens the outward heat flow, creating a positive feedback loop that drives the system into runaway gravothermal collapse, as described in \citet{Balberg2002}.

The dynamical evolution of SIDM halos---encompassing both core formation and gravothermal collapse---has been widely explored as a potential explanation for a range of astrophysical phenomena. In particular, SIDM-driven heat transport and collapse provide a natural mechanism for the observed diversity of galactic rotation curves, in which halos of similar mass exhibit a wide range of central densities \citep[e.g.,][]{Oman2015, Kamada2017, Ren2018, Nadler2023, Roberts2025}. In the context of strong gravitational lensing, the extremely dense cores produced during deep gravothermal collapse have been proposed as a viable explanation for the compact dark perturbers inferred from lensing observations \citep[e.g.,][]{Nadler2023, Li2025, Tajalli2025, Kong2025, Yu2025, Simona2026}; similar ideas have also been explored for dense perturbers of stellar streams \citep[e.g.,][]{Zhang2025}. Moreover, the rapid deepening of the gravitational potential during the collapse phase has also been suggested as a possible catalyst for the formation of massive quasars and black holes at high redshift \citep[e.g.,][]{Balberg2002, Pollack2015, Feng2022, Meshveliani2023, Feng2025, Jiang2025}.

A robust test of SIDM predictions requires accurate modeling of SIDM halos across their full gravothermal evolution. To meet this challenge, a range of complementary approaches have been developed. $N$-body simulations \citep{Burkert2000, Kochanek2000, Koda&Shapiro2011, Vogelsberger2012, Rocha2013, Fry2015, Banerjee2020, Correa2022, Fischer2025, Nadler2025, Bosch_2026} provide a direct description of SIDM dynamics across different stages of halo evolution by following particle trajectories and self-interactions, but their computational cost limits systematic exploration of large parameter spaces and long evolutionary timescales. Recently, symmetry-exploiting approaches have been developed that retain a first-principles description of gravitational dynamics while achieving substantial computational speedups \citep[e.g.,][]{Kamionkowski2025, Gurian2025}. 

An alternative description is provided by gravothermal fluid models \citep{Balberg2002, Koda&Shapiro2011, Pollack2015, Essig2019, Nishikawa2020, Feng2021, Gu2026}, which treat SIDM halos as conducting, self-gravitating fluids and evolve coupled equations for mass, momentum, and heat transport, providing a practical balance between computational efficiency and a physically motivated, approximate description of SIDM gravothermal evolution. More recently, parametric models calibrated to SIDM simulations have been developed to efficiently describe halo evolution across different gravothermal phases. These models have been further refined by incorporating gravothermal fluid results to improve their accuracy in specific regimes \citep{YangDN2024, YangDN2025, Hou2025}.

Another widely used semianalytic approach is the isothermal Jeans model \citep{Kaplinghat2014, Kaplinghat2016}, which provides a computationally efficient framework for modeling SIDM halos. By assuming approximate thermal equilibrium in the inner halo, the dark matter density profile can be obtained directly from the Jeans and Poisson equations, enabling rapid exploration of SIDM parameter space. The validity of this equilibrium-based approach has been tested in cosmological settings \cite[e.g.,][]{Robertson2021, Ren2018}, where the Jeans model was shown to reproduce the spherically averaged density profiles of SIDM halos in cosmological simulations across a wide range of halo masses. The framework has also been generalized to describe elliptical halo configurations by \citet{Bautista2025}. However, the applicability of the Jeans model during the highly dynamical core-collapse phase remains uncertain. \citet{Jiang2023} showed that the matching between the inner isothermal solution and the outer collisionless halo admits both low- and high-density branches, and \citet{YangSQ2024} proposed a mirroring prescription to associate the high-density branch with the collapse phase. \citet{Jia2026} further refined the numerical implementation of this framework and extended its application to modeling SIDM halo structures. While this extension shows encouraging agreement with gravothermal fluid models, a systematic validation against accurate, first-principles simulations at sufficiently high resolution is still required.

In this work, we use a state-of-the-art, high-resolution isolated $N$-body simulation \citep{Fischer2025} to test the reliability of the isothermal Jeans model in describing the full gravothermal evolution of SIDM halos, including the core-collapse phase modeled via the mirroring of the high-density solution. By performing time-resolved comparisons of three-dimensional density and velocity-dispersion profiles, as well as projected lensing observables, we demonstrate that the Jeans model provides a fast and broadly reliable framework for modeling the gravothermal evolution of SIDM halos, including the collapse phase.

The remainder of this paper is organized as follows. Section~\ref{sec:method} presents our methodology, including a review of the isothermal Jeans model (Section \ref{sec:method_IJMoverview}), specifications of the idealized isolated $N$-body simulation used for validation (Section \ref{sec:method_nbody}), and the procedures for computing the benchmark quantities used in the comparison (Section \ref{sec:method_quantity}). Section~\ref{sec:results} provides a comprehensive comparison between the Jeans model and $N$-body simulation, examining both the evolutionary tracks in lensing parameter space (Section \ref{sec:results_lensing}) and the internal density and velocity-dispersion profiles (Section \ref{sec:results_profiles}). A discussion of their implications follows in Section~\ref{sec:discussion}. Finally, we summarize our conclusions in Section~\ref{sec:conclusions}, and additional material is provided in the Appendices.

\section{Method} \label{sec:method}
In this section, we first briefly overview the isothermal Jeans model (Section~\ref{sec:method_IJMoverview}), then describe the $N$-body simulation used for comparison (Section~\ref{sec:method_nbody}), and finally detail the calculation of the quantities used for our benchmark test (Section~\ref{sec:method_quantity}).

\subsection{Overview of isothermal Jeans model} \label{sec:method_IJMoverview}
The isothermal Jeans model, as originally introduced in \citet{Kaplinghat2014, Kaplinghat2016}, is built on the basic assumption that self-interactions in the dense, central region of a SIDM halo lead to the formation of an isothermal core. The structure of this core is determined by solving the coupled Jeans and Poisson equations. Assuming a pure dark matter halo with isotropic velocities, the density profile of the inner, isothermal core is uniquely determined by two parameters: its central density ($\rho_0$) and its constant velocity dispersion ($\nu_0$). In the halo's outer regions, where self-interactions are negligible, the density is assumed to follow a standard Navarro--Frenk--White (NFW; \citealt{Navarro1996}) profile. The inner isothermal core and the outer NFW halo are smoothly joined at a stitching radius, $r_1$. This radius marks the boundary where the local scattering rate, assuming a constant total cross section per particle mass ($\sigma_m$), is high enough for a particle to undergo one interaction, on average, over time ($t$). \citet{Kaplinghat2016} originally defined $t$ as the halo age, approximately 10 Gyr for galaxies at $z\sim0$. In this work, we instead interpret $t$ as the effective evolution time since the onset of self-interactions. In practice, we treat the product $t\,\sigma_m$ as the evolutionary variable. By solving for $r_1$ at increasing values of this product, we obtain the full evolutionary sequence of the SIDM halo within the Jeans framework. To match the setup of \citet{Fischer2025}, we adopt $\sigma_m = 80\,\mathrm{cm^2\,g^{-1}}$ and compute the corresponding gravothermal evolution of the density profile.

To extend the Jeans framework into the core-collapse regime, we adopt the mirroring prescription proposed by \citet{YangSQ2024}, which exploits the two branches of matched isothermal solutions: a low-density branch and a high-density branch that approach each other at a merger time, $t_{\rm merge}$. The merger time $t_{\rm merge}$ serves as a practical indicator of the transition to the gravothermal core-collapse regime in the isothermal Jeans framework \citep{Jiang2023}. The high-density branch is then mirrored about $t_{\rm merge}$ to provide a continuous mapping onto the subsequent collapse evolution, enabling the construction of SIDM halo profiles up to the latest collapse stages accessible within this framework. Since the accuracy of this extension during collapse is not guaranteed \textit{a priori}, a central goal of this work is to benchmark its predictions against a high-resolution, first-principles $N$-body simulation. The numerical procedure for locating both solution branches and determining $t_{\rm merge}$ is based on the method introduced by \citet{Jiang2023}, with the specific implementation adopted here described in \citet{Jia2026}.

\subsection{The reference \texorpdfstring{$N$}{N}-body simulation} \label{sec:method_nbody}
This reference simulation from \citet{Fischer2025} was performed with the \texttt{OpenGadget3} code \citep{Dolag2026} and is based on the SIDM implementation introduced by \citet{Fischer2021, Fischer2022, Fischer2024}. The initial halo density profile is described by an NFW profile with a characteristic density $\rho_s = 4.42\times10^{7}\,\mathrm{M_\odot\,kpc^{-3}}$ and a scale radius $r_s = 1.28\,\mathrm{kpc}$. Using a Hubble constant of $H_0 = 67.36\,{\rm km\,s^{-1}\,Mpc^{-1}}$ \citep{PlanckCollaboration2020} to define the critical density at $z=0$, these parameters correspond to a halo mass of $M_{200}=2.56\times10^9\,{\rm M_\odot}$ and a concentration of $c_{200}=22.54$, which we adopt in our isothermal Jeans model for comparison. In the simulation, the halo is represented by $5\times10^7$ particles, yielding a particle mass of $\sim 50\,\mathrm{M_\odot}$. It assumes isotropic, velocity-independent scattering with a total cross section per physical dark matter particle mass of $\sigma_m = 80 \text{ cm}^2\text{g}^{-1}$ and the gravitational softening length is $\epsilon = 1.0$ pc. Other key numerical choices include a kernel size defined by the 48 nearest neighbors ($N_\mathrm{ngb}=48$) and a gravitational time-step parameter of $\eta = 2.5 \times 10^{-2}$. The latter controls the accuracy of the adaptive time-stepping scheme ($\Delta t \propto \sqrt{\eta}$) together with an additional time-step criterion for the self-interactions, which is critical for resolving the rapid dynamical evolution within the deepening potential well. \citet{Fischer2025} demonstrated that with this setup the simulation is robust, and its convergence has been verified well into the gravothermal core-collapse phase, making it an ideal reference for our study. The simulation produces snapshots at intervals of $\Delta t\approx0.1$~Gyr over a total duration of $\sim5$~Gyr.

\subsection{Calculating the quantities used for comparison}\label{sec:method_quantity}
Using the halo parameters ($M_{200}$, $c_{200}$) and the self-interaction cross section ($\sigma_m$) specified in Section~\ref{sec:method_nbody}, we employ the Jeans model and obtain the  corresponding density profile as a function of evolution time. We track the halo's evolution from 0.001 Gyr up to the formal end point of the Jeans model, $t_{\rm coll} \equiv 2 \times t_{\rm merge}$, which marks the validity limit of the mirroring strategy. Over this interval, we sample the central density, $\rho_0$, and velocity dispersion, $\nu_0$, at 1500 logarithmically spaced time steps to construct robust interpolation functions, $\rho_0(t)$ and $\nu_0(t)$. Since these two parameters uniquely define the inner isothermal core, this allows us to reconstruct the full density profile, $\rho(r)$, at any evolutionary stage. To capture the initial state correctly, we manually assign the NFW profile at $t=0$. 

We derive the radial velocity-dispersion squared, $\sigma_r^2(r)$, by solving the spherical Jeans equation using the modeled density and enclosed-mass profiles,
\begin{flalign}\label{eq:velocity_dispersion}
\sigma_r^2(r) =
\frac{\mathrm{G}}{r^{2\beta}\rho(r)}
\int_r^{\infty}
\frac{r'^{2\beta}\rho(r')M(r')}{r'^2}\,dr', &&
\end{flalign}
where $\rm{G}$ is the gravitational constant, $M(r)$ denotes the mass enclosed within spherical radius $r$, and we assume an isotropic velocity distribution by setting the anisotropy parameter $\beta=0$.

The projected enclosed mass, $M_{\rm 2D}$, and projected logarithmic density slope, $\gamma_{\rm 2D}$, are critical observables in strong gravitational lensing and serve as powerful probes of dark matter physics. Therefore, beyond the 3D density profiles, we place emphasis on benchmarking the evolution of these projected properties against the simulation. To ensure fair comparison, we calculate these quantities using the same method as in the simulation analysis in \citet{Fischer2025}; see details in Appendix~\ref{sec:appendixA1}.

\section{Comparison of Isothermal Jeans model and \texorpdfstring{$N$-body}{N-body} simulation}\label{sec:results}
In this section, we present a comprehensive comparison between the results of our semianalytical isothermal Jeans model and the high-resolution isolated $N$-body simulation from \citet{Fischer2025}. We organize the comparison into two parts: first, we evaluate the model's predictions for projected lensing observables (Section~\ref{sec:results_lensing}); second, we investigate the internal density and velocity-dispersion profiles (Section~\ref{sec:results_profiles}). This allows us to verify the model's consistency with simulation during the core-formation phase while explicitly identifying the deviations that characterize the late-stage core collapse.

\subsection{Evolution in the lensing parameter space}\label{sec:results_lensing}
\begin{figure}
    \centering
	\includegraphics[width=\columnwidth]{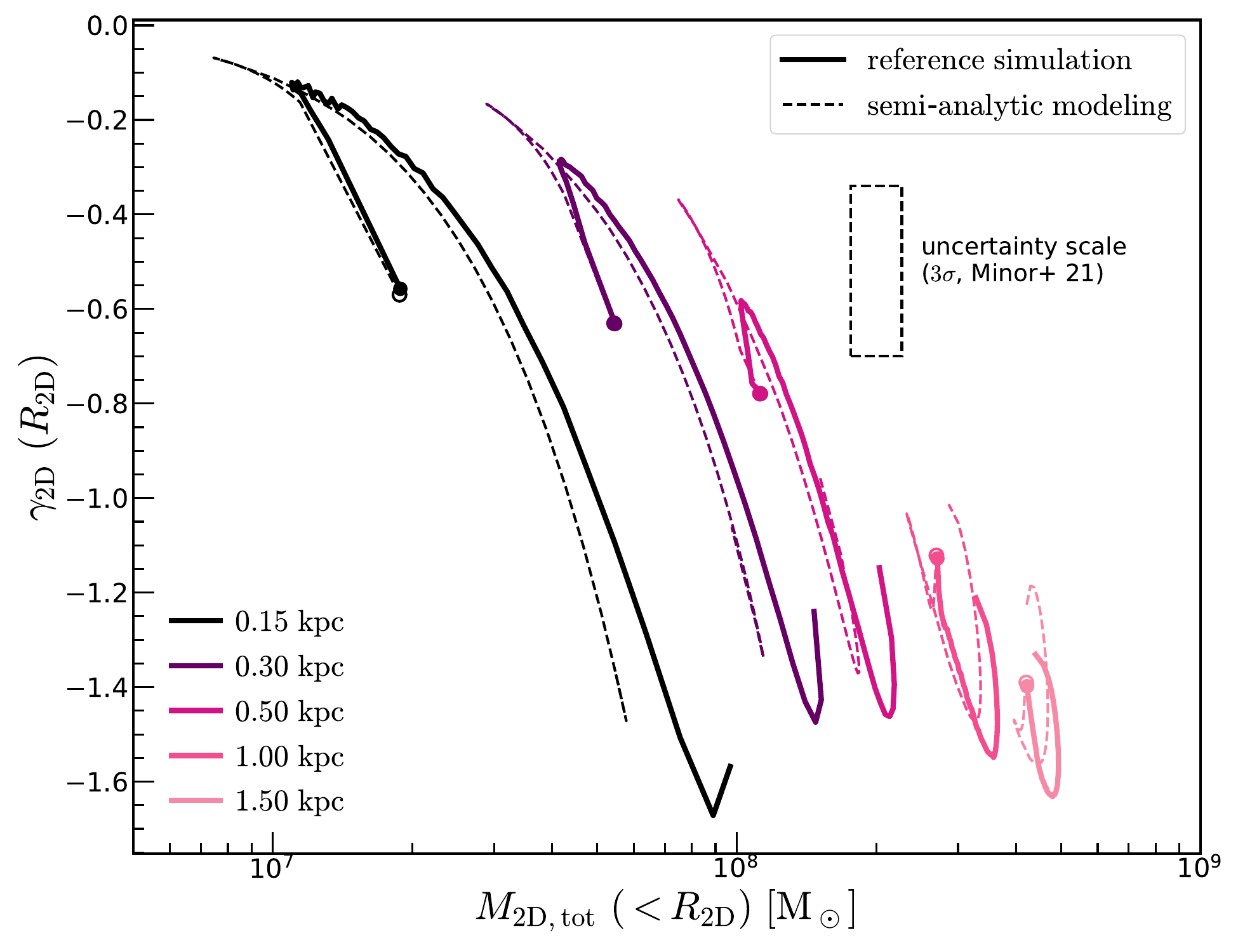}
    \caption{Evolutionary tracks of the projected logarithmic density slope ($\gamma_{\rm 2D}$) and enclosed mass ($M_{\rm 2D}$) calculated at various projected radii ($R_{\rm 2D}$). Solid and dashed lines represent the results from the reference $N$-body simulation and the isothermal Jeans model, respectively. The dots mark the starting point of each track ($t=0$), corresponding to the initial NFW profile. The black dashed box represents the $3\sigma$ observational uncertainty from \citet{Minor2021}, shown for scale only.}
    \label{Figure1}
\end{figure}

In Fig.~\ref{Figure1}, we display the evolutionary tracks of the lensing properties---specifically the projected logarithmic density slope, $\gamma_{\rm 2D}$, versus the enclosed mass, $M_{\rm 2D}$---measured at various projected radii ranging from 0.15 kpc to 1.5 kpc. The starting points of the evolutionary tracks ($t=0$), marked by dots, correspond to the projected properties of the initial NFW halo. For the initial halo configuration, both the simulation and the Jeans model yield nearly identical results, although the former is based on live particles. This validates that our method for calculating $M_{\rm 2D}$ and $\gamma_{\rm 2D}$ is consistent with that adopted in \citet{Fischer2025}. Therefore, any discrepancies discussed below arise from the underlying difference between the Jeans model and the $N$-body simulation.

\begin{figure*}
    \centering
	\includegraphics[width=\textwidth]{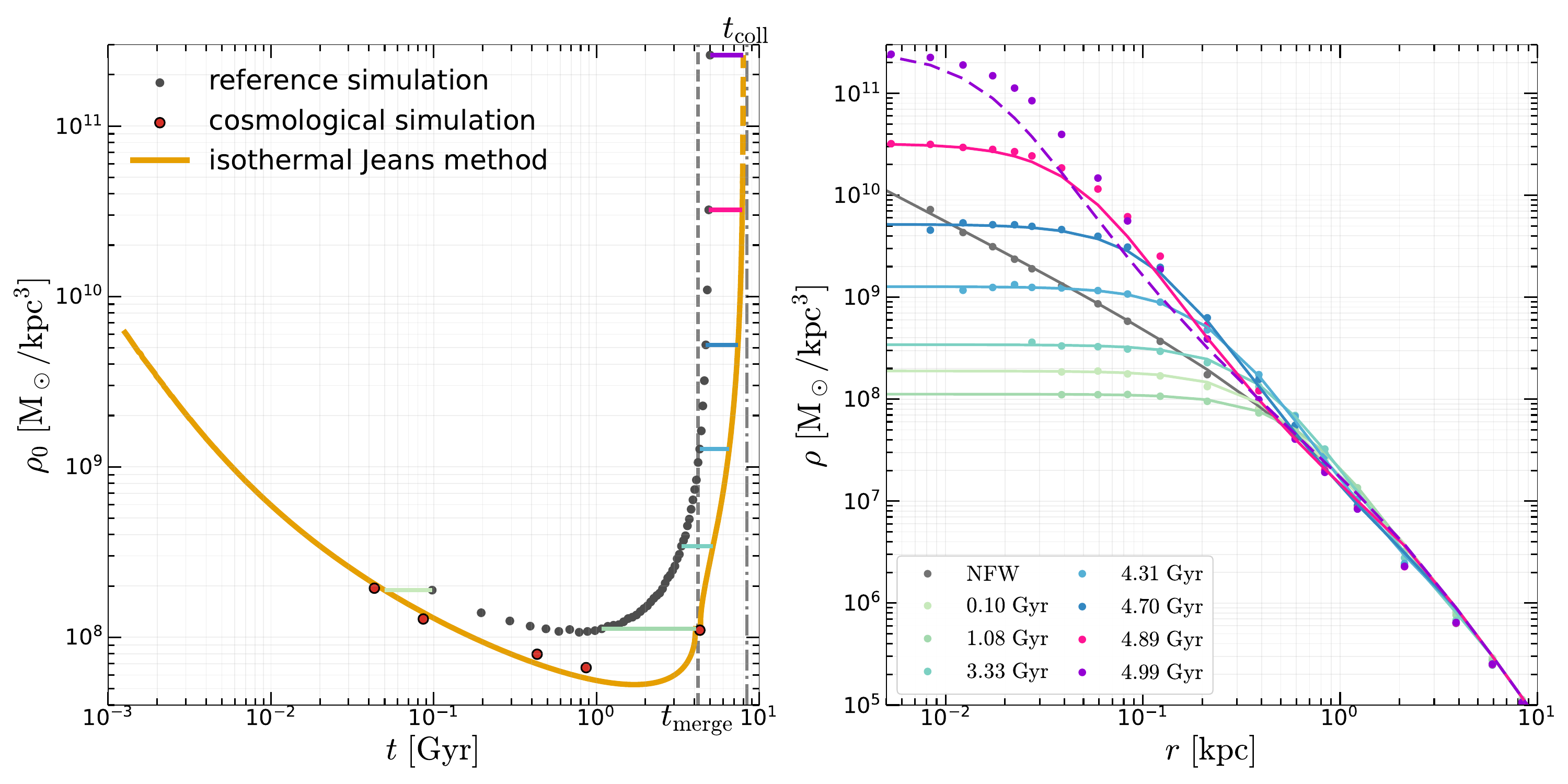}
    \caption{Comparison of the temporal evolution of the central density and the radial structure. Left: central density $\rho_0$ as a function of time $t$ for the isolated reference simulation (gray points, showing all available simulation snapshots), the rescaled cosmological Pippin simulation (red circles), and the isothermal Jeans model (orange curve). The dashed segment of the orange curve indicates a regime where isothermal solutions do not exist and is constructed using extrapolated values of $\rho_0$. The gray vertical lines indicate the merger time $t_{\rm merge}$ and the collapse time $t_{\rm coll}$  within the Jeans framework, while the colored horizontal bars mark the epochs at which density profiles are shown in the right panel. Right: spherically averaged density profiles at the selected times. Dots show the reference simulation results, where only radial bins with particle numbers exceeding 200 are included. Curves show the corresponding isothermal Jeans profiles. The dashed segment corresponds to the profile constructed using extrapolated $\rho_0$ and $\nu_0$. Colors indicate the snapshot times as labeled in the legend, and the gray curve denotes the initial CDM NFW halo.}
    \label{Figure2}
\end{figure*}

The evolution exhibits a clear two-phase behavior. Starting from the initial NFW profile, the projected slope rapidly flattens while the enclosed mass decreases, reflecting the core-formation phase. The reference simulation produces snapshots at intervals of $\Delta t\approx0.1$~Gyr. As a result, the core-formation stage is sampled by only a limited number of snapshots, leading to an apparently straight segment at the beginning of the simulation track in Fig.~\ref{Figure1}. In contrast, the isothermal Jeans model is not subject to this temporal sampling limitation and resolves a much smoother evolutionary track. Following this stage, the projected slope steepens and the enclosed mass increases, marking the transition into gravothermal core collapse. In the simulation, at $R_{\rm 2D}=0.15$~kpc, $\gamma_{\rm 2D}$ evolves to approximately $-1.7$ and $M_{\rm 2D}$ increases to $\sim10^8~{\rm M_\odot}$, corresponding to a factor of $\sim5$ enhancement relative to the initial halo. Note that in the Jeans model, the central region does not evolve to become as dense as in the simulation, due to the limitations that we will discuss later.

At the late stage of the collapse, the evolutionary tracks develop a clear turnover, which becomes more pronounced at larger projected radii. This turnover arises because the steepest part of the density profile shifts progressively inward during deep core collapse, while the projected slope $\gamma_{\rm 2D}$ is evaluated at a fixed radius. As a result, the slope measured at that radius no longer steepens monotonically and instead reverses its trend. In addition, the enclosed mass within fixed radii exhibits a slight late-time decrease, which first appears at larger radii (see also Fig.~9 of \citealt{Fischer2025}).

Overall, the evolutionary tracks from the two methods agree well, with the exception of two notable discrepancies. The first appears at the point of maximum core size, where the Jeans model tends to predict a more extended core than the simulation. The second discrepancy arises during the core-collapse phase, when the Jeans model fails to achieve the same level of central concentration as the simulation, leading to a lower enclosed mass and a shallower projected density slope in the inner regions. Notably, these differences diminish with increasing projected radius, since self-interactions play a significant role only in the halo’s central region.

To illustrate the scale of the model deviations relative to observational uncertainty, we overplot the observational uncertainty scale inferred from \citet{Minor2021} for the SDSS~J0946+1006 subhalo. \citet{Li2025} showed that SIDM halos with $M_{200}\sim 5\times10^{10}$--$10^{11}\,{\rm M_\odot}$ can reproduce the observed properties of this system. In units of the scale radius, 1~kpc in those halos corresponds to $\sim 0.1$--$0.4\,r_s$, mapping to physical radii of $\sim 0.15$--$0.5$~kpc in the simulated halo studied here. At these radii, the discrepancies between the simulation and the semianalytic predictions during core collapse remain smaller than the observational uncertainties for this system. This suggests that, for strong-lensing analyses at comparable mass scales and observational precision, the limitations of the Jeans model are unlikely to dominate the inferred constraints.

\subsection{Density and velocity-dispersion profiles}\label{sec:results_profiles}
To identify the source of the deviations in the projected lensing observables, we compare the underlying three-dimensional density profiles. We adopt the core central density, $\rho_0$, as the reference quantity for comparison. To extract $\rho_0$ from each simulation snapshot while minimizing numerical fluctuations, we fit the spherically averaged density profile within the inner 0.5~kpc using the King model, restricting the fit to radial bins that each contain more than 200 particles,
\begin{flalign}
\rho(r) = \rho_0 \left[ 1 + \left( \frac{r}{r_c} \right)^2 \right]^{-3/2}. &&
\end{flalign}
$\rho_0$ is the fitted central density, and $r_c$ is a characteristic core radius. As demonstrated by \citet{Fischer2025}, the King model provides a reasonable description of the inner structure of SIDM halos over the collapse stages resolved in current simulations; see also \citet{Zhang2025} for an application to dense, core-collapse SIDM halos.

In the left panel of Fig.~\ref{Figure2}, we show the evolution of the central density as a function of time, comparing the reference $N$-body simulation with the prediction from the isothermal Jeans model. Regarding the evolutionary timescale, the onset of core collapse occurs about two times later in the Jeans model than in the reference simulation, leading to a more extended core prior to collapse. This behavior is consistent with the discrepancies identified in Fig.~\ref{Figure1}, where the Jeans model yields a less concentrated mass distribution prior to the collapse phase. Notably, the minimum central density predicted by the Jeans model is approximately half that of the simulation. 

During the collapse phase, the Jeans model predicts states of equal central density at times that are systematically later by approximately 0.2 dex in $t$ compared to the simulation. Since the gravothermal timescale scales inversely with the self-interaction cross section, this offset would translate into an overestimation of the inferred cross section by approximately a factor of 1.6 if the Jeans model were used to infer the cross section from a given collapse time. However, as we will discuss in Section~\ref{sec:discussion_timescale}, the collapse timescale predicted by the Jeans model can be better aligned with that found in cosmological SIDM simulations.

In the right panel of Fig.~\ref{Figure2}, we compare the density profiles of the simulation and the Jeans model to examine the halo structure in more detail. We select a set of snapshots in the simulation whose central densities span the full dynamic range of the SIDM evolutionary sequence. For each selected snapshot, we identify the corresponding one in the Jeans model by matching the value of $\rho_0$, as indicated by the horizontal markers in the left panel. Over most of the evolutionary history, the density profiles predicted by the two methods agree remarkably well. Noticeable differences emerge only in the late stages of core collapse, when $\rho_0 \gtrsim 200\,\rho_s$. In this regime, the halo predicted by the Jeans model develops a less concentrated central region than the simulated halo. This mass deficit manifests as a shallower projected logarithmic slope and a lower enclosed mass, consistent with the deviations previously shown in Fig.~\ref{Figure1}.

The emergence of these late-time discrepancies coincides with a breakdown in the practical numerical construction of the isothermal solutions of the Jeans model. At the latest evolutionary stages accessible to the Jeans model---defined by the end point $t_{\rm coll} \equiv 2\,t_{\rm merge}$---the adopted procedure no longer yields a self-consistent high-density branch that can be smoothly matched to the outer NFW profile, and hence no solution suitable for the mirroring prescription can be identified. For this simulated halo, self-consistent isothermal solutions become unavailable approximately $0.5$~Gyr (about 6\% of $t_{\rm coll}$) before this end point. Over this interval, the central density $\rho_0(t)$ and velocity-dispersion parameter $\nu_0(t)$ are extrapolated in time, and the resulting density profiles should accordingly be regarded as approximate estimates rather than genuine model predictions. The extrapolated Jeans model results are presented by the dashed curves in Fig.~\ref{Figure2}, and the final snapshot of the reference simulation falls within this regime.

\begin{figure}
    \centering
	\includegraphics[width=\columnwidth]{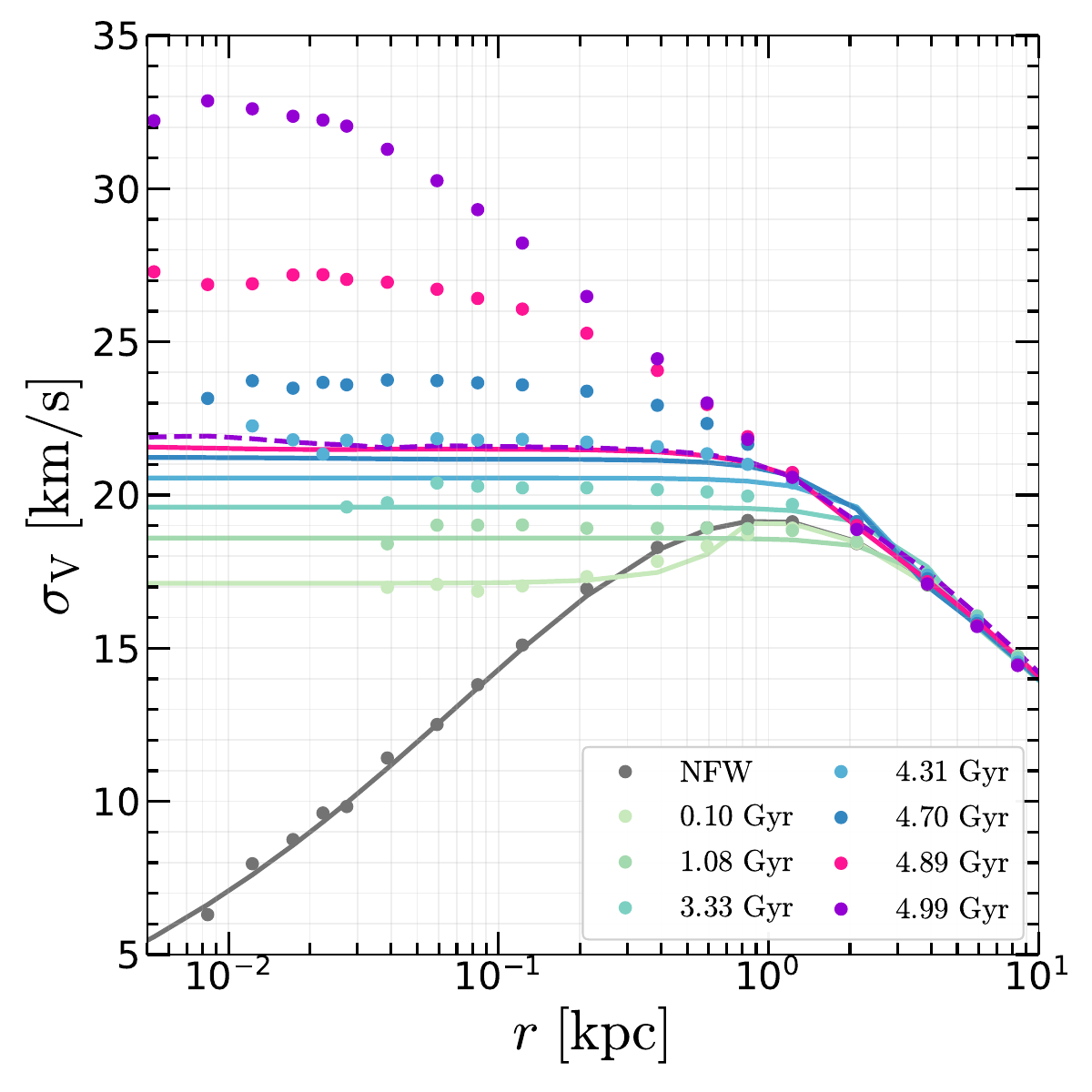}
    \caption{Radial profiles of the one-dimensional velocity dispersion, $\sigma_{\rm v}(r)$, for the SIDM halo at selected times. Dots show the spherically averaged results of the reference simulation, where only radial bins with particle numbers exceeding 200 are included. Curves represent the corresponding isothermal Jeans model results. The dashed segment indicates the profile computed using extrapolated values of $\rho_0$ and $\nu_0$ in the absence of isothermal solutions. Colors indicate the snapshot times as labeled in the legend, and the gray curve denotes the initial CDM NFW halo.}
    \label{Figure3}
\end{figure}

In Fig.~\ref{Figure3}, we compare the velocity-dispersion profiles for the same selected snapshots and model stages as in the density-profile comparison shown in the right panel of Fig.~\ref{Figure2}. The isothermal Jeans model reasonably captures the broad radial shape of these profiles up to approximately $4\,{\rm Gyr}$. However, a modest normalization offset in the inner region is already visible at the $1.08\,{\rm Gyr}$ snapshot, around maximum core expansion, and persists thereafter, with the simulated profiles lying systematically above the Jeans model predictions.

The discrepancy becomes much more severe after $t\gtrsim 4\,{\rm Gyr}$. In this phase, the simulated halo develops a rapidly increasing central velocity dispersion and a more compact approximately isothermal region, whereas the Jeans model maintains a more extended matched isothermal core. Consequently, the model increasingly underestimates the velocity-dispersion normalization and fails to capture the late-time radial evolution of the profile. The largest normalization discrepancy reaches $\sim 10\,{\rm km\,s^{-1}}$ at the final simulation snapshot.

Although the density profiles in the right panel of Fig.~\ref{Figure2} appear broadly consistent in the direct log-log comparison, this does not necessarily imply that the reconstructed velocity-dispersion profiles should agree at the same level. In the isotropic Jeans reconstruction of Eq.~\eqref{eq:velocity_dispersion}, $\sigma_r^2(r)$ depends on the full radial structure through the non-local integral involving $\rho(r')M(<r')/r'^2$, as well as on the local density normalization $1/\rho(r)$. Therefore, residual differences in the detailed density and enclosed-mass structure can lead to more visible differences in the reconstructed velocity-dispersion profiles, especially during the late core-collapse phase. We have also verified that the assumption of velocity isotropy is not the primary source of this discrepancy.

To further clarify the origin of this late-time velocity-dispersion discrepancy, we compare the central velocity-dispersion normalization of the simulation with that of the corresponding Jeans model solutions. For each simulation snapshot, we identify the Jeans model solution with the same central density $\rho_0$ and compare the corresponding values of $\nu_0$. This comparison shows that the Jeans model $\nu_0$ remains close to the simulation value before $\sim 4\,{\rm Gyr}$, but becomes substantially lower during the late core-collapse phase. Thus, even when the late-time snapshots are matched by central density, the Jeans model increasingly underestimates the central velocity-dispersion normalization. This underestimation corresponds to the growing late-time normalization offset seen in Fig.~\ref{Figure3}. Further details of the comparison of $\nu_0$ are provided in Appendix~\ref{sec:appendixA2}.

\section{Discussion} \label{sec:discussion}
In this section, we compare the evolutionary timescales predicted by the isothermal Jeans model with those from the cosmological simulations of \citet{Elbert2015} (Section~\ref{sec:discussion_timescale}). We then extend our analysis to include the parametric model proposed in \citet{YangDN2024, YangDN2025}, which provides a computationally efficient framework for modeling SIDM evolution (Section~\ref{sec:discussion_parametric}). These complementary comparisons allow us to assess the applicability and limitations of the Jeans model in a more comprehensive manner.

\subsection{Interpretation of the Timescale Discrepancy} \label{sec:discussion_timescale}
As shown in the left panel of Fig.~\ref{Figure2}, the timescale for core collapse appears to be somewhat longer in the Jeans model than in the isolated simulation. It is therefore interesting to further compare these results with cosmological SIDM simulations, in which ongoing mass accretion may influence the thermal evolution of the halo and potentially delay the onset of gravothermal collapse. Here, we take the Pippin halos from the cosmological zoom-in simulations of \citet{Elbert2015} as an example for comparison, in which the same halo is evolved to the same cosmic time ($z=0$) with different self-interaction cross sections. Following the approach of \citet{Jiang2023}, we infer a dimensionless gravothermal evolutionary track of the central density, $\tilde{\rho}_0(\tilde t)$, from the cosmological Pippin simulations, which can then be rescaled to the physical density and time for the halo studied in this work. This mapping exploits the approximate self-similarity of SIDM gravothermal evolution in the long-mean-free-path regime (LMFP), in which the self-interaction cross section sets the characteristic gravothermal timescale \citep[e.g.][]{Balberg2002, Nishikawa2020, Essig2019}.

We adopt the NFW scale density $\rho_s$ and scale radius $r_s$ inferred from the CDM counterpart of the Pippin halos. The central density measured from each SIDM simulation is then normalized as $\tilde{\rho}_0 = \rho_0 / \rho_s$. The characteristic gravothermal timescale $t_0$ scales as $t_0 \propto \sigma_m^{-1} r_s^{-1} \rho_s^{-3/2}$ \citep{Essig2019}, and the corresponding dimensionless time is given by $\tilde t = t/t_0$, with $t$ fixed to the cosmic age at $z=0$ (13.7~Gyr). Under this mapping, simulations with different self-interaction cross sections correspond to different locations along a single gravothermal evolutionary track in the $\tilde{\rho}_0-\tilde t$ plane. Finally, by rescaling the dimensionless variables using $r_s$, $\rho_s$, and $t_0$ of the halo considered in this work, we obtain the evolutionary tracks of the Pippin simulations shown in the left panel of Fig.~\ref{Figure2}.

We find that the rescaled cosmological results follow the isothermal Jeans trajectory more closely than the isolated $N$-body simulation. As speculated in \citet{Jiang2023}, the Jeans model exhibits closer agreement with cosmological results because it adopts the NFW profile as a boundary condition rather than as an initial condition. More generally, the agreement of semianalytic SIDM models with cosmological simulations can also depend on the calibration of characteristic interaction scales, for example through the matching prescription at $r_1$ \citep[e.g.][]{Ren2018}.

As a result, the evolution parameter $t$ carries different interpretations in different frameworks. In isolated $N$-body and gravothermal fluid models, an NFW halo is evolved forward for a physical duration $t$. In contrast, the isothermal Jeans model fixes the outer NFW structure and uses $t$ to parameterize the depth of the self-interaction-driven modification in the central region. In this sense, $t$ in the Jeans framework represents the effective age of an SIDM halo whose core has evolved while the outer halo remains anchored to an NFW-like envelope. Accordingly, the timescale discrepancy between the isolated $N$-body simulation and the isothermal Jeans model should not be interpreted as a failure of the semianalytic approach, but rather as a reflection of the more rapid gravothermal evolution expected in isolated systems, where environmental effects that may delay collapse, such as mergers, are absent. A quantitative mapping between cosmological accretion or merger histories and the gravothermal evolutionary timescale inferred by the Jeans framework is left to future work, for example by exploiting cosmological SIDM zoom-in simulation suites spanning a wide range of halo masses, such as the Concerto SIDM suite \citep{Nadler2025}.

\subsection{Comparison with Parametric Gravothermal SIDM Model} \label{sec:discussion_parametric}
A complementary fast approach to model the evolution of SIDM halos is provided by the parametric gravothermal framework, developed in \citet{YangDN2024,YangDN2025} and subsequently extended by \citet{Hou2025}. This approach is motivated by theoretical studies showing that, in the LMFP regime, the gravothermal evolution of SIDM halos is approximately universal, with halos evolving through a sequence of quasi-equilibrium states whose density and velocity-dispersion profiles remain well captured throughout the core-formation phase \citep[e.g.][]{Outmezguine2023, Zhong2023, GadNasr2024}. As a result, once the evolution is expressed in terms of the dimensionless gravothermal phase, $\tilde t \equiv t/t_0$ (denoted as $\tau$ in \citealt{Hou2025}), halos with different masses and self-interaction cross sections follow nearly identical evolutionary trajectories, allowing their density profiles to be represented by a single, parameterized family of profiles (see Fig.~1 and Fig.~2 of \citealt{Hou2025}).

Therefore, the parametric model describes the full gravothermal evolution with an analytic density profile whose parameters are given by fitting functions calibrated to high-resolution SIDM simulations. For a given halo, the model requires only the initial NFW parameters and the SIDM cross section to determine the collapse time; the subsequent structure is obtained by selecting the appropriate gravothermal phase, $\tilde t$. Owing to this simplicity and computational speed, the parametric approach is well suited for statistical applications, such as population-level lensing predictions.

\begin{figure}
    \centering
	\includegraphics[width=\columnwidth]{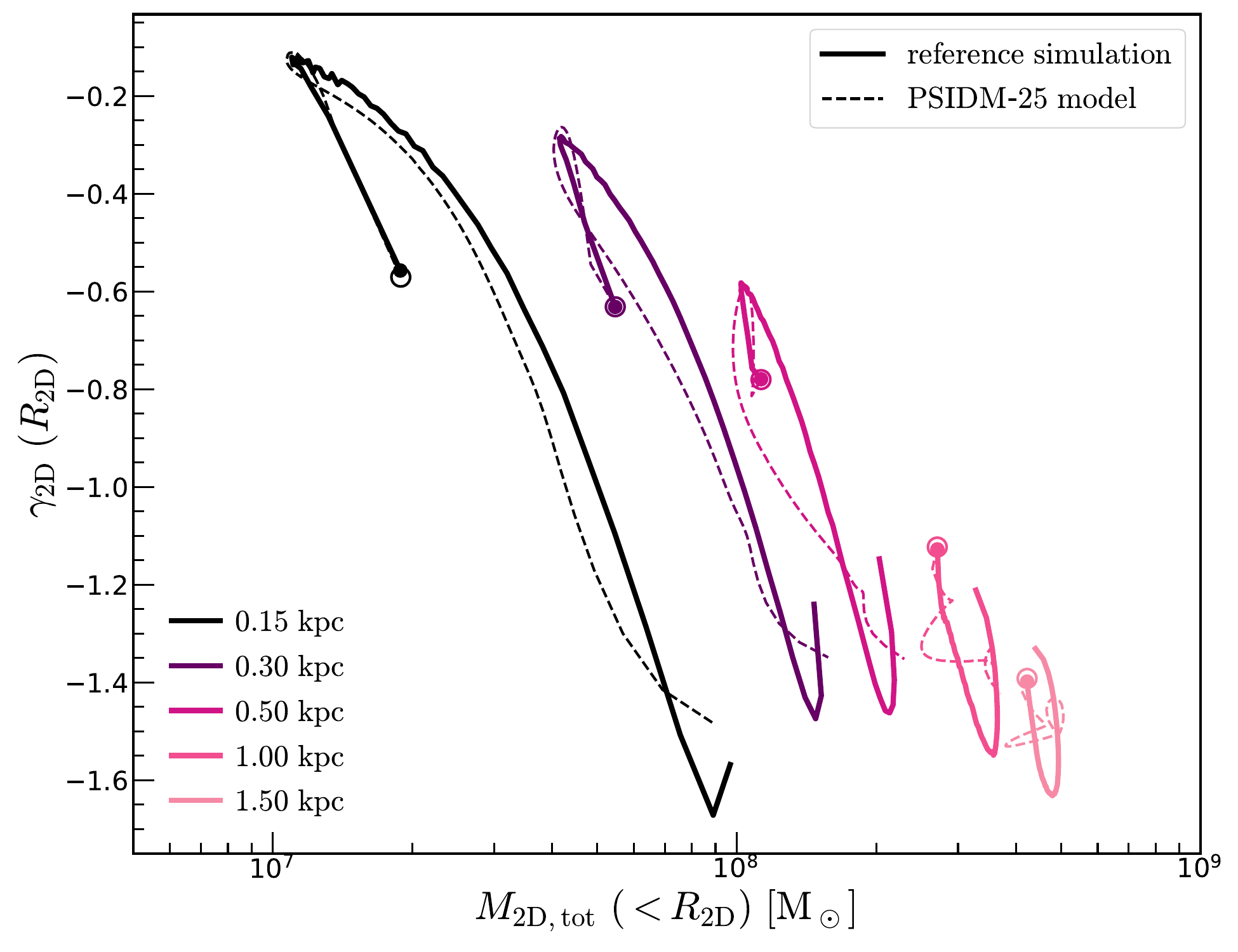}
    \caption{Same as Fig.~\ref{Figure1}, but comparing the results from the isolated $N$-body simulation with the parametric SIDM model based on the gravothermal fluid solution from \citet{Hou2025}.}
    \label{Figure4}
\end{figure}

In Fig.~\ref{Figure4}, we show the evolution of the projected mass $M_{\rm 2D}$ and slope $\gamma_{\rm 2D}$ predicted by the parametric model, compared with the isolated $N$-body simulation. The agreement is particularly good at small projected radii (e.g. $R_{\rm 2D} \lesssim 0.3\,\mathrm{kpc}$), where the parametric trajectories closely follow the $N$-body results throughout the evolution. More generally, during the core-formation phase, the parametric model provides a good description of the evolutionary trends across all projected radii considered here. Deviations become more significant at larger projected radii during the deep core-collapse phase, where the parametric model develops additional structure in the $M_{\rm 2D}$–$\gamma_{\rm 2D}$ plane that is not present in the simulation. These deviations likely reflect the limited flexibility of the analytic parameterization to capture the rapid structural evolution in this late stage.

\section{Conclusions} \label{sec:conclusions}
In this work, we evaluated the performance of the isothermal Jeans model by comparing its predictions with results from an isolated $N$-body simulation. We find that:
\begin{itemize}
    \item The Jeans model accurately reproduces the projected lensing observables of the simulation throughout the gravothermal evolution, with deviations remaining well within current observational uncertainties at comparable mass scales.

    \item The Jeans model reproduces the three-dimensional density profiles of the isolated simulation with excellent accuracy over most of the halo evolution, while noticeable deviations emerge only in the deep core-collapse phase.

    \item The Jeans model predicts a more extended core and a delayed collapse time relative to the isolated simulation. This slower evolution is consistent with cosmological Pippin SIDM simulations and reflects the NFW outer boundary adopted in the Jeans framework.
    
    \item The Jeans model captures the broad radial shape of the velocity-dispersion profiles through core formation and into the early stages of collapse, although it does not fully capture the rapid rise in central velocity dispersion and the shrinking approximately isothermal region during the later collapse phase.

\end{itemize}
In summary, we have demonstrated that the isothermal Jeans model, extended to the core-collapse phase, provides an efficient and practical tool for modeling the evolution of SIDM halos, with broad potential applications. Further comparison with cosmological zoom-in SIDM simulations across a range of halo masses (e.g., \citealt{Nadler2025}) would provide a more comprehensive assessment of its performance. In addition, exploration of baryonic effects in realistic cosmological environments would further refine its applicability. We leave these extensions to future work.

\begin{acknowledgments}
This project was initiated at the Valencia SIDM Workshop 2025. S.L. and R.L. acknowledge the support of the National Natural Science Foundation of China (No. 12588202), the science research grants from the China Manned Space Program (Nos. CMS-CSST-2025-A03 and CMS-CSST-2025-A20), and the CAS Project for Young Scientists in Basic Research (No. YSBR-062). M.S.F. acknowledges the support of the Alexander von Humboldt Foundation through a Feodor Lynen Research Fellowship. Z.J. and F.J. acknowledge the support of the National Natural Science Foundation of China (No.~12473007) and the Beijing Natural Science Foundation (QY23018). H.B.Y. acknowledges the support of the U.S. Department of Energy (DE-SC0008541) and the John Templeton Foundation (\#63599). The opinions expressed in this publication are those of the authors and do not necessarily reflect the views of the funding agencies. The analysis and plotting scripts used in this work, together with the processed data products needed to reproduce the figures, are publicly available at \url{https://github.com/Shubo143/Jeans_vs_Nbody}.
\end{acknowledgments}

\appendix
\twocolumngrid
\section{Calculating the lensing properties} \label{sec:appendixA1}
To enable a direct comparison with the $N$-body simulation results, we compute the projected quantities using the same numerical procedures adopted in the simulation analysis. The projected mass $M_{\rm 2D}$ is calculated as the mass within a cylinder of radius $R_{\rm 2D}$. For the density profile $\rho(r)$ in our model, we first compute the surface density profile $\Sigma(R)$ by integrating along the $z$-axis, with the integration limit set to $z_{\rm lim}=3 \times r_{200}$:
\begin{flalign}
    \Sigma(R)=\int^{z_{\rm lim}}_{-z_{\rm lim}}\rho(\sqrt{R^2+z^2})\,\mathrm{d}z. &&
\end{flalign}
The projected mass $M_{\rm 2D}(< R_{\rm 2D})$ is then obtained by integrating $\Sigma(R)$ from 0.001 kpc to $R_{\rm 2D}$.

For the projected logarithmic density slope, $\gamma_{\rm 2D}$, we employ a finite-difference approximation consistent with the binning scheme used in the simulation. At each radius $R_{\rm 2D}$, we define two adjacent logarithmic annuli sharing $R_{\rm 2D}$ as their common boundary: the inner annulus spans $[R_{\rm low}, R_{\rm 2D}]$, and the outer annulus spans $[R_{\rm 2D}, R_{\rm high}]$. We compute the average surface density in each annulus, $\Sigma_{\rm inner}$ and $\Sigma_{\rm outer}$, by dividing the total surface mass in the annulus by its area. The local logarithmic slope at $R_{\rm 2D}$ is then approximated as
\begin{flalign}
    \gamma_{\rm 2D} =
    \frac{\ln(\Sigma_{\rm outer}) - \ln(\Sigma_{\rm inner})}
    {\frac{1}{2}[\ln(R_{\rm high}) - \ln(R_{\rm low})]}, &&
\end{flalign}
where the denominator corresponds to the logarithmic width of a single annulus, since $\ln(R_{\rm high}) - \ln(R_{\rm low})$ represents the total width of the two adjacent bins combined.

\setcounter{figure}{0}
\renewcommand{\thefigure}{B\arabic{figure}}
\begin{figure}
    \centering
	\includegraphics[width=\columnwidth]{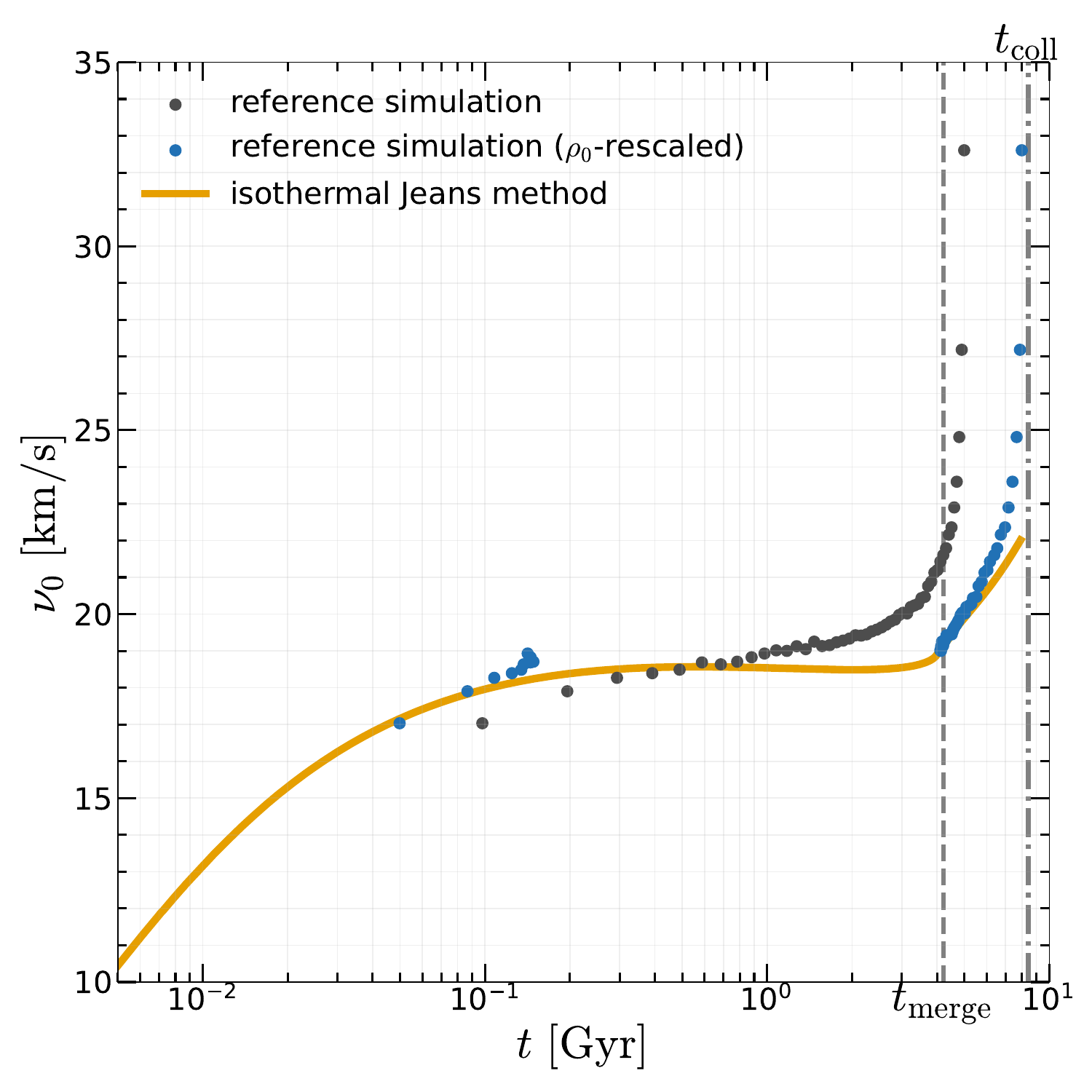}
    \caption{Comparison of the temporal evolution of the central velocity dispersion. Gray points show $\nu_0$ measured directly from the reference $N$-body simulation. Blue points show the same simulation values of $\nu_0$, but plotted at the $\rho_0$-matched Jeans model time, i.e. the epoch at which the Jeans model has the same central density as the corresponding simulation snapshot. The orange curve shows the prediction of the isothermal Jeans model. The gray vertical lines indicate the merger time $t_{\rm merge}$ and the collapse time $t_{\rm coll}$ within the Jeans framework.}
    \label{FigureB1}
\end{figure}

\begin{figure*}
    \centering
	\includegraphics[width=\textwidth]{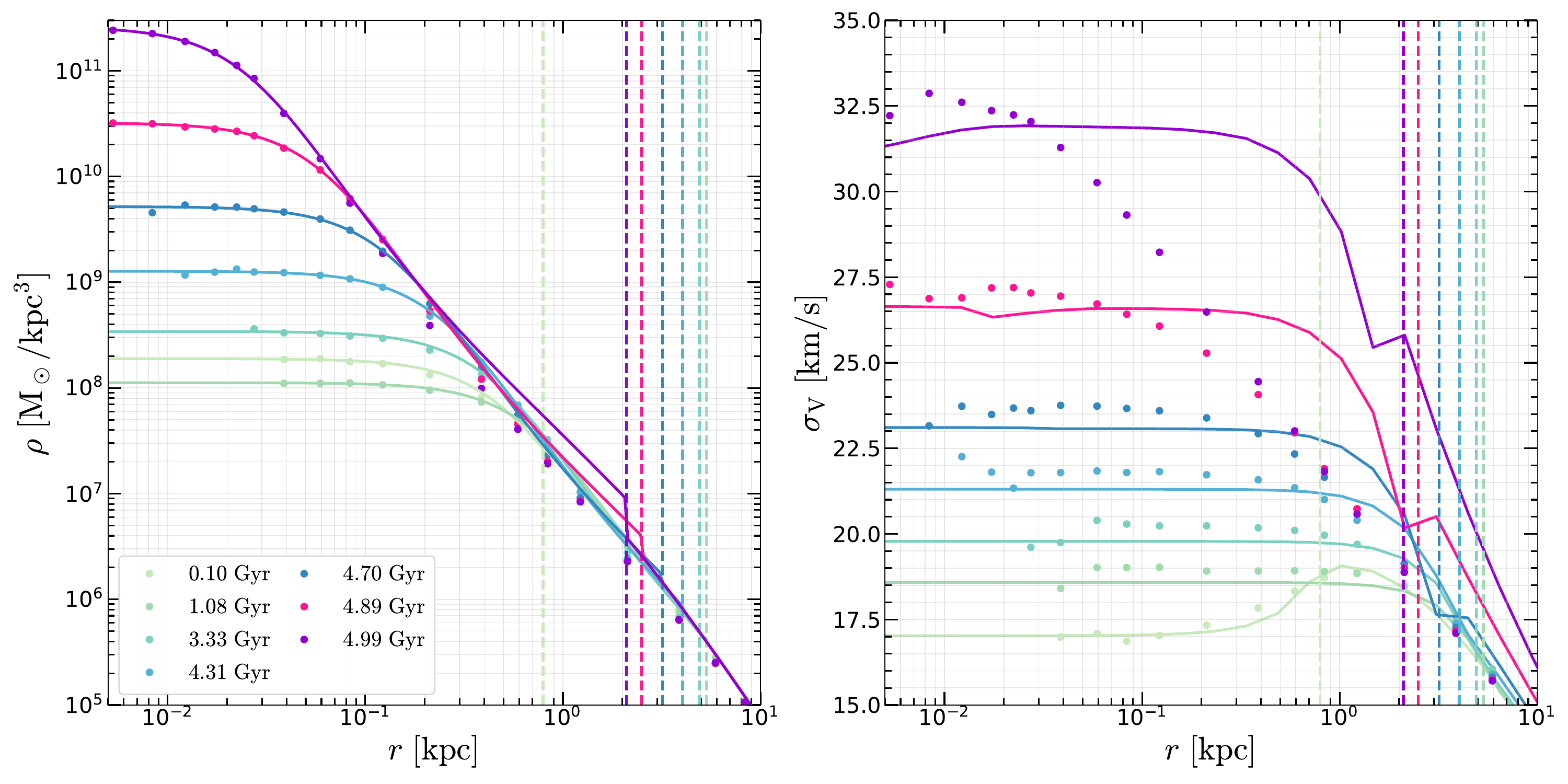}
    \caption{Radial profiles of the density (left) and one-dimensional velocity dispersion (right) for the diagnostic Jeans construction in which the central velocity-dispersion parameter $\nu_0$ is set to the value measured from the reference $N$-body simulation. Dots show the spherically averaged simulation profiles at the selected epochs, where only radial bins with particle numbers exceeding 200 are included. Curves represent the constructed Jeans model results. Colors indicate the snapshot times as labeled in the legend, with vertical lines of the same colors marking the corresponding stitching radii in the Jeans model.}
    \label{FigureB2}
\end{figure*}

\section{Impact of central velocity dispersion on model solutions} \label{sec:appendixA2}
In Fig.~\ref{FigureB1}, we compare the temporal evolution of the central velocity dispersion, $\nu_0$, between the reference $N$-body simulation and the isothermal Jeans model to investigate the origin of the discrepancies in the velocity-dispersion profiles during the core-collapse phase. For each simulation snapshot, $\nu_0$ is measured as the median velocity dispersion of the five innermost radial bins that each contain at least 200 particles. As discussed in Section~\ref{sec:discussion_timescale}, and illustrated in Fig.~\ref{Figure2}, the evolutionary timescales of the isolated $N$-body simulation and the Jeans model are not directly comparable. To enable a meaningful comparison of the central velocity dispersion, for each simulation snapshot we identify the isothermal solution with the same central density $\rho_0$ and reset the time coordinate to the corresponding epoch in the Jeans model. The resulting central velocity dispersion $\nu_0$ measured from the simulation is then plotted in Fig.~\ref{FigureB1} at this matched time, enabling a direct comparison of $\nu_0$ at equivalent evolutionary stages.

As shown in Fig.~\ref{FigureB1}, although both the simulation and the Jeans model exhibit an increasing central velocity dispersion during the core-collapse phase, the Jeans model fails to capture the rapid rise of $\nu_0$ seen in the simulation. As a result, a discrepancy exceeding $\sim10~{\rm km\,s^{-1}}$ accumulates by the final simulation snapshot. This deviation reflects a structural limitation of the isothermal Jeans model: in the current implementation, the region inside the stitching radius $r_1$ is enforced to follow a single isothermal solution smoothly matched to an outer NFW profile. However, the velocity-dispersion profiles in Fig.~\ref{Figure3} show that the spatial extent of the isothermal core in the simulation shrinks rapidly during the core-collapse phase, whereas the stitching radius $r_1$ in the Jeans model does not contract at a comparable rate. Consequently, the model overestimates the size of the isothermal region and underestimates the true central velocity dispersion $\nu_0$.

To isolate the impact of this underestimated \(\nu_0\), we perform a diagnostic construction shown in Fig.~\ref{FigureB2}. In Figs.~\ref{Figure2} and \ref{Figure3}, for each simulation snapshot $i$, we identify the Jeans model epoch $t_i^{\rm J}$ by matching the central density, $\rho_0^{\rm J}\left(t_i^{\rm J}\right)=\rho_{0,i}^{\rm sim}$, and then use the corresponding self-consistent Jeans model parameters $\{\rho_0^{\rm J}(t_i^{\rm J}),\,\nu_0^{\rm J}(t_i^{\rm J}),\,r_1^{\rm J}(t_i^{\rm J})\}$ to construct the density and velocity-dispersion profiles. In Fig.~\ref{FigureB2}, we keep the same $\rho_0$-matched epoch $t_i^{\rm J}$, but replace the Jeans model central velocity-dispersion parameter with the value measured from the simulation, $\nu_0^{\rm J}(t_i^{\rm J}) \rightarrow \nu_{0,i}^{\rm sim}$. The diagnostic profile is therefore constructed from $\{\rho_0^{\rm J}(t_i^{\rm J}),\,\nu_{0,i}^{\rm sim},\,r_1^{\rm J}(t_i^{\rm J})\}$, with $\rho_0^{\rm J}(t_i^{\rm J})$ and $r_1^{\rm J}(t_i^{\rm J})$ kept fixed from the original $\rho_0$-matched Jeans solution. Because $\nu_0$ is imposed by hand in this construction, the resulting isothermal profile is not guaranteed to join smoothly onto the outer NFW profile, particularly during the core-collapse phase. We plot this constructed density profile in the left panel of Fig.~\ref{FigureB2}. In the right panel, we show the corresponding velocity-dispersion profile reconstructed from the density and enclosed-mass profiles in the left panel using Eq.~\eqref{eq:velocity_dispersion}, and compare both panels with the simulation results. The stitching radius at each epoch is also indicated by a vertical line.

Ignoring the distortions near $r_1$ introduced by the discontinuity in the stitched density profile, we find that imposing the simulation value of $\nu_0$ leads to a substantially improved agreement between the inner density profiles of the Jeans model and the simulation, even for the final snapshot in the core-collapse phase. In this diagnostic construction, the remaining discrepancy in the velocity-dispersion profiles is primarily associated with the Jeans model overestimating the spatial extent of the isothermal core. A small residual offset in the central normalization still persists even after $\nu_0$ is matched to the simulation value, indicating that part of the discrepancy is associated with the idealized spherical, isotropic Jeans reconstruction of Eq.~\eqref{eq:velocity_dispersion} itself, rather than with the value of $\nu_0$ alone. Incorporating a more realistic, time-dependent treatment of the isothermal core size during gravothermal collapse therefore represents a key direction for future improvements of the isothermal Jeans framework. 

\bibliography{ref}{}
\bibliographystyle{aasjournal}

\end{document}